\documentclass[naturemag,article,amsmath,amssymb]{revtex4}

\usepackage{graphicx}

\usepackage{graphicx}
\usepackage{xcolor}

\begin{document}

\title{Non-thermal transport of energy driven by photoexcited carriers in switchable solid states of GeTe}  


\author{R. Gu$^{1}$, T. Perrault$^{1}$, V. Juv\'e $^{1}$, G. Vaudel$^{1}$, M. Weis $^{1}$, A. Bulou $^{1}$, N. Chigarev$^{2}$, A. Levchuk$^{1}$, S. Raetz$^{2}$, V. E.  Gusev$^{2}$, Z. Cheng $^{3}$, H. Bhaskaran$^{3}$\footnote{ Electronic address: harish.bhaskaran@materials.ox.ac.uk}, P. Ruello$^{1}$\footnote{ Electronic address: pascal.ruello@univ-lemans.fr}}

\affiliation{
$^{1}$Institut des Mol\'ecules et Mat\'eriaux du Mans, UMR 6283 CNRS, Le Mans Universit\'e, 72085 Le Mans,  France\\
$^{2}$Laboratoire d'Acoustique de Le Mans Universit\'e, UMR CNRS 6613, Le Mans Universit\'e, 72085 Le Mans, France\\
$^{3}$ Department of Materials, University of Oxford, United Kingdom.
}

\begin{abstract}

Phase change alloys have seen widespread use from rewritable optical discs to the present day interest in their use in emerging neuromorphic computing architectures. In spite of this enormous commercial interest, the physics of carriers in these materials is still not fully understood. Here, we describe the time and space dependence of the coupling between photoexcited carriers and the lattice in both the amorphous and crystalline states of one phase change material, GeTe. We study this using a time-resolved optical technique called picosecond acoustic method to investigate the \textit{in situ} thermally assisted amorphous to crystalline phase transformation in GeTe. Our work reveals a clear evolution of the electron-phonon coupling during the phase transformation as the spectra of photoexcited acoustic phonons in the amorphous ($a$-GeTe) and crystalline ($\alpha$-GeTe) phases are different. In particular and surprisingly, our analysis of the photoinduced acoustic pulse duration in crystalline GeTe suggests that a part of the energy deposited during the photoexcitation process takes place over a distance that clearly exceeds that defined by the pump light skin depth. In the opposite, the lattice photoexcitation process remains localized within that skin depth in the amorphous state. We then demonstrate that this is due to supersonic diffusion of photoexcited electron-hole plasma in the crystalline state. Consequently these findings prove the existence of a non-thermal transport of energy which is much faster than lattice heat diffusion. 

\end{abstract}

\maketitle

\section{Introduction} Phase change materials (PCM) have become a critical aspect of advanced computing architectures~\cite{gho}, ranging from arithmetic operations~\cite{wri,hos,fel1}, neuromorphic computing\cite{gong,jafari,cheng} and more recently in matrix-vector multiplications for machine learning and artificial intelligence applications~\cite{fel2,legallo,rios}. In spite of their large deployment in industry and products, and a wealth of studies~\cite{wuttig,kolobov,wuttig2}, the physics of the dynamics of phase transformation is still under intense study~\cite{raoux}. These technologies are underpinned by a reversible phase transformation from an amorphous to crystalline phase thermally, optically or electrically assisted~\cite{wuttig,kolobov,wuttig2,raoux}. The need for fast writing-erasing processes in emerging GHz-THz information technologies will benefit from ultrafast photoinduced transformation. However, this is presently not possible, and mastering this technology requires a better understanding of the electron and phonon dynamics at short time and space scales in non-equilibrium states, e.g. after a photoexcitation process~\cite{kolobov,wuttig2,raoux,rios2,kolobovPRB}. While phase transformation was initially described as a thermally assisted process, the recent debate about the crucial role of photoexcited carriers in the amorphization process ($a$-GeTe to $\alpha$-GeTe) has questioned a potentially complex non-thermal process \cite{kolobov,kolobovPRB,hada} which calls for clarifications.  In this photoinduced transformation process, the photoexcited carriers as suspected to modify the inter-atomic potential in a non-thermal way, leading in some circumstances to the lattice instability~\cite{kolobov,kolobovPRB,hada}. The photoexcited carrier-lattice coupling is at the core of the crucial mechanism of transformation of the light energy into the lattice energy and represents the necessary step for phase transformation. Contradictory assumptions of the characteristic spatial extension of the coupling between photoexcited carriers and phonon currently exist. It has been considered that the photoexcited carrier-lattice coupling in PCM is restricted to the optical skin depth (light penetration)~\cite{kolobov,kolobovPRB,hada,hase,walde,zalden}, while another report suggests possible ballistic-electrons effect~\cite{fons}, without any clear and direct conclusion about this crucial physical phenomenon. Using a time-resolved optical method~\cite{tom1,gusev1993,ruelloLU}, we demonstrate that photo-carriers supersonically diffuse in the crystalline phase $\alpha$-GeTe over a distance which is around 7 times the optical skin depth, i.e. the photoexcited carriers transport energy via a non-thermal process around 7 times deeper than the depth of light absorption. However, we show that this supersonic electron-hole plasma expansion is not effective in the amorphous state a-GeTe. 
This technique called picosecond acoustic method enables depth profiling of the coupling between photoexcited carriers and the lattice. Such a method has already been used to show how photoexcited carriers rapidly diffuse non-thermally out of the optical skin in GaAs~\cite{wright,kent} or Ge~\cite{chigarev} semiconductors. It has also been employed to reveal ultrafast non-local heating in metals~\cite{tas,lejmanJOSA,ruello2015}. Time-resolved optical techniques, such as coherent optical phonons spectroscopy, have already been used to probe the amorphous to crystalline phase transformation, but this method is only sensitive to the optical phonons (i.e. unit cell) and cannot map the photoexcited carriers diffusion~\cite{dekorsy,boschetto}. Photoinduced strain studies have already been reported in GST compounds (Ge$_2$Sb$_2$Te$_5$)~\cite{hada,hase,lindenberg}. Some differences in the acousto-optic signals have been reported~\cite{hase,lindenberg} but this non-thermal supersonic electronic diffusion contribution to the acoustic pulse broadening was not discussed or observed. The crucial parameter in this study, besides the high sensitivity of the optical measurements, is the choice of the PCM thin film thickness. Most of previous studies~\cite{kolobovPRB,hada,hase,walde,zalden,fons} concerned thin films whose thicknesses were typically the same order of the optical pump light penetration depth ($\sim$ 30nm) in the crystalline PCM. These thin thicknesses were usually chosen for necessary experimental constraint associated to time-resolved X-ray or electron-diffraction~\cite{kolobovPRB,hada,walde,zalden,fons}. For such thin layers, there is a natural in-depth confinement of the photoexcited electron-hole plasma preventing the electron-hole plasma to expand. In this work, we have then performed in situ pump-probe experiments on a much thicker film with a GeTe film having a thickness of 380nm to let the electron-hole plasma to expand over a distance of around 200-400 nm as we will show. We then evidence an ultrafast non-thermal transport of energy  that it is much faster than the ballistic acoustic phonon propagation which sets a limit of the phonon thermal transport.

\begin{figure*}[t!]
\centerline{\includegraphics[width=16cm]{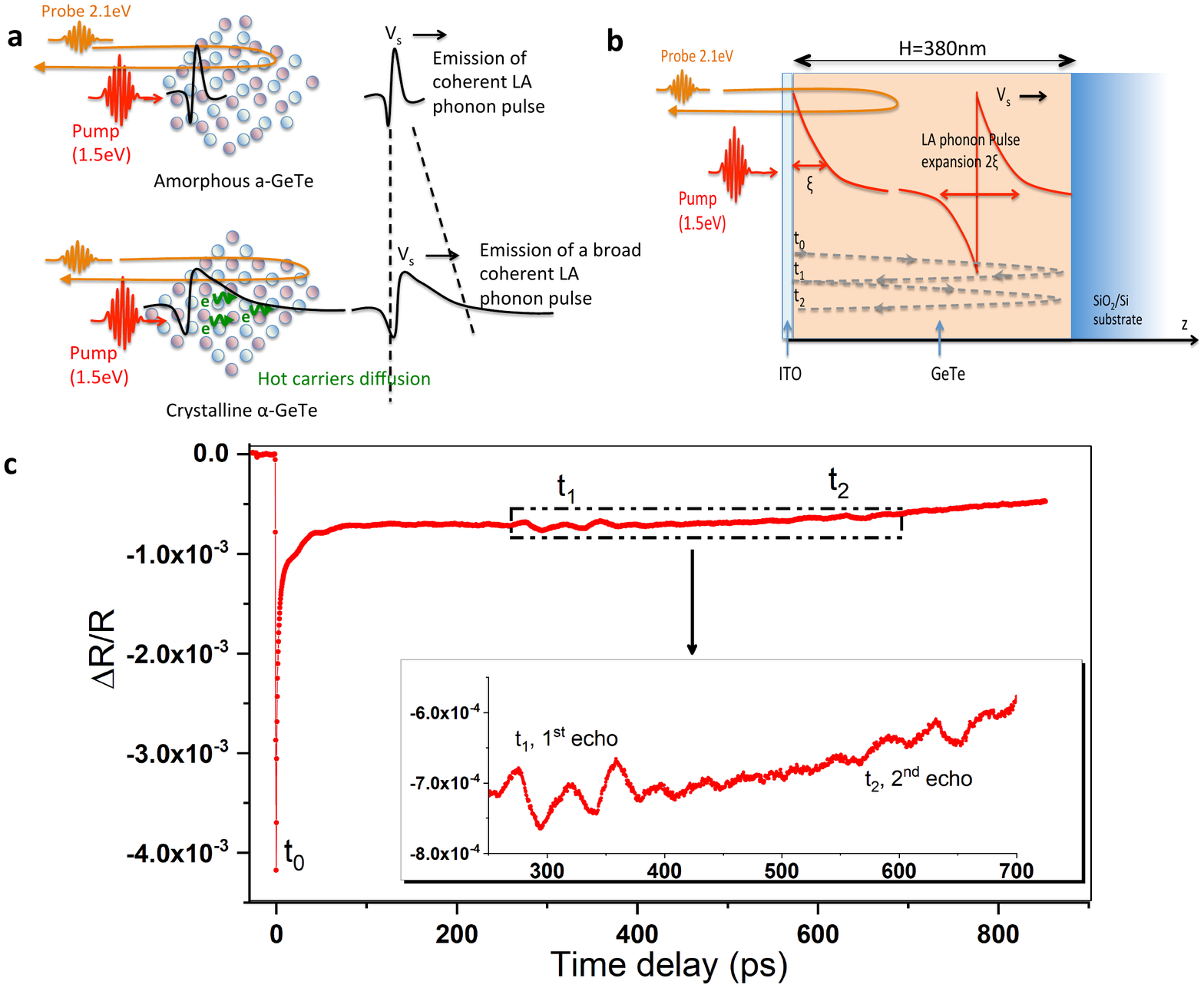}}
\caption{\label{fig1} 
Sketch of the photoexcitation of GeTe in the amorphous and crystalline phases where the photoinduced acoustic phonon pulses have strong differences related to the electronic properties of the material. (b) Description of the 380nm thick thin GeTe film studied with time-resolved optical measurement. The femtosecond laser pump pulse generates a strain pulse in the near surface that travels back and forth in the film depicted by the arrows. These acoustic pulses are monitored in the time-domain thanks to a delayed probe pulse. (c) Typical transient optical reflectivity signals obtained with a pump power of 5 mW in the amorphous phase revealing clearly two acoustic echoes.}
\end{figure*}

\begin{figure*}[t!]
\centerline{\includegraphics[width=14cm]{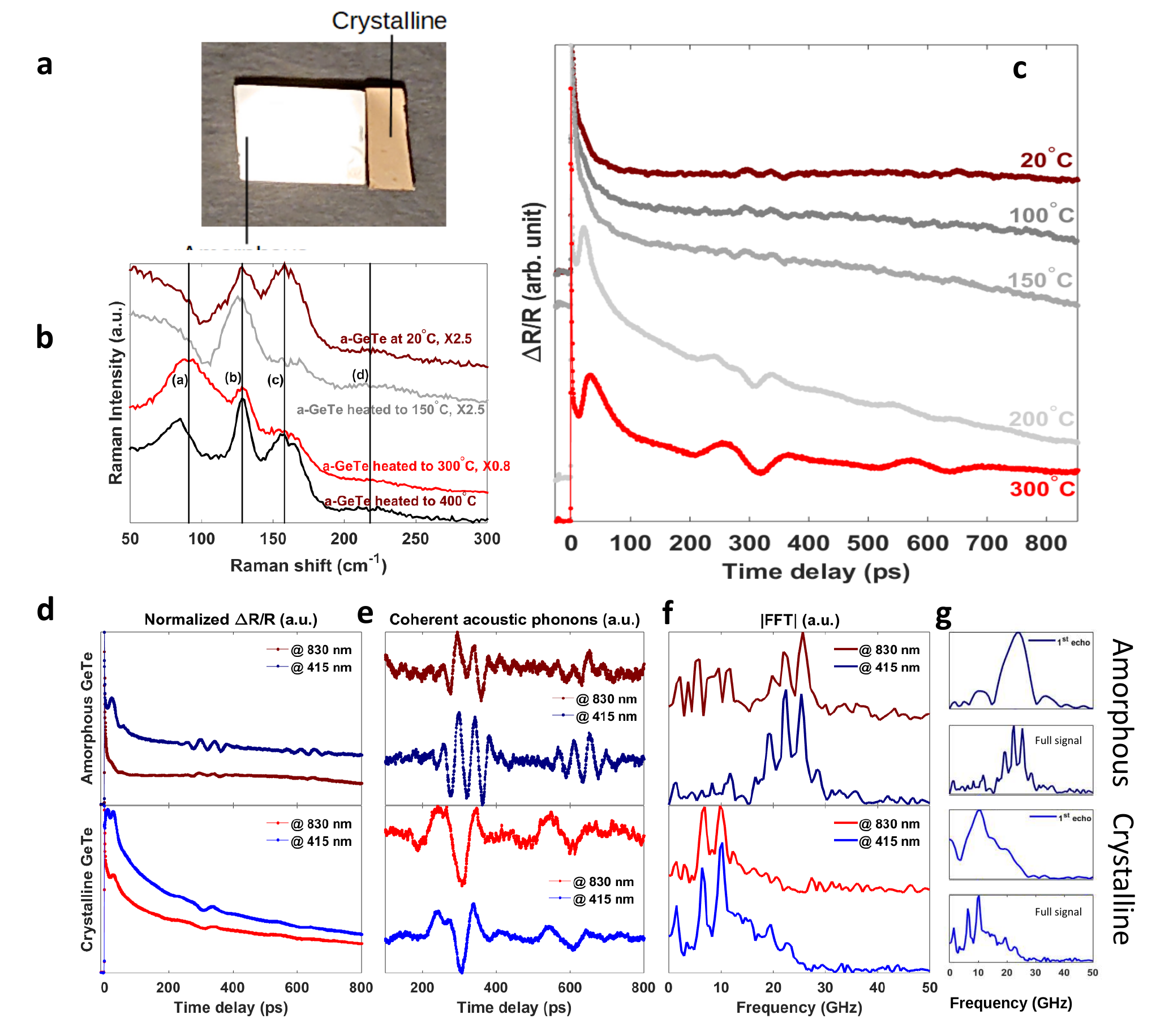}}
\caption{\label{fig2} 
(a) Image with a white light source showing the optical contrast between the amorphous and crystalline GeTe. (b) Raman spectra recorded as a function of the temperature. (c) Transient optical reflectivity signals versus the temperature during the amorphous to crystalline phase transformation (pump @830nm). (d) Comparison of room temperature transient optical reflectivity signals obtained with two pump photon energies (@830nm-1.5eV and @ 415nm-3 eV) for both the amorphous and crystalline phase. (e) Extracted acoustic signal. (f) Fast Fourier transform (FFT) of the acoustic signals shown in (e). (g) FFT of the first acoustic echo compared to the FFT of the full signal.}
\end{figure*}

\section{Experimental methods}
A principle of the pump-probe experiments described in this article is shown in Fig. \ref{fig1}a where the light-matter interaction leads to the generation of the longitudinal acoustic phonon (LA) whose spectrum (pulse duration) provides information on the photoexcited carrier dynamics, in particular on the hot carriers diffusion. A description of the nanostructure under investigation is shown in Fig. \ref{fig1}b. The GeTe sample has been grown with physical deposition method (RF sputtering). The sample has a thickness of around 380 nm ($\pm 10$ nm). It is covered by a protecting thin (10nm) ITO transparent layer. GeTe stoichiometry is 50/50 and is deposited on Si substrate with a SiO$_2$ buffer layer.  The pump-probe method is based on a Ti:sapphire femtosecond laser that delivers nJ pulses of around 200fs of duration. The repetition rate is 76MHz. The experiments were conducted with a two-color pump-probe scheme with a pump laser radiation centered at 830nm (1.5eV) or at its second harmonic 415 nm (3eV) obtained by the optical second harmonic generation (SHG) in a BBO crystal. The probe beam wavelength was always fixed at 583nm (2.2eV). The focusing diameter of the pump and probe beams were $\sim$ 14$\mu$m and $\sim$ 8$\mu$m. The pump and probe are focused on the GeTe surface with a quasi normal incidence. The 830nm is the harmonic of a Ti:sapphire oscillator while the probe beam wavelength is controlled thanks to an Optical Parametric Oscillator (OPO) 
For picosecond acoustics experiments versus temperature, a Linkam furnace was used to realized \textit{in situ} temperature dependence measurements. \\
In this experiment, the light energy is absorbed and distributed over a characteristic distance which leads to the generation of a strain pulse. The latter one travels back and forth (after reflection on the substrate) in the layer and periodic acoustic echoes are detected at times t$_1$ and t$_2$ as indicated in Fig. \ref{fig1}c. Our goal is to probe how the energy absorbed by the electronic subsystem after the femtosecond excitation is transferred to the phonon subsystem via the electron-phonon coupling in the amorphous and crystalline states. The investigations are then conducted within the perturbative approach, i.e. we aim at investigating the physical properties change under a thermally assisted phase transformation and not under light-induced phase transformation. As a consequence, careful measurements were done by selecting the proper pump and probe fluences to prevent any transformation of GeTe under the light excitation. It is known indeed that photoassisted excitation process can transform the amorphous phase into the crystalline phase and vice-versa~\cite{kolobovPRB,siegel}. To ensure that we do not induce the amorphous-crystalline phase transformation (and the reverse one) under the action of the pump beam, we have calibrated our measurements as described in Supplementary Note S1~\cite{SM}. We have shown that working with pump fluence ranging from 2.5 to 5 mW, i.e. 60 $\mu$J/cm$^{-2}$ (for our laser repetition rate of 80 MHz) allows us to remain under the optical threshold (see Supplementary Figures 4 and 5~\cite{SM}). 

\section{Results}
A typical transient optical reflectivity signal is shown in Fig. \ref{fig1}c for the case of amorphous GeTe. We identify first a sharp variation just after the pump excitation (time t$_0$) corresponding to the excitation of the electronic subsystem followed by a rapid electron-hole-phonon thermalization. Slight modulation of the transient optical reflectivity signal are observed in the time range ($\sim$0-50 ps) and more clearly pronounced oscillations at around $\sim$300 ps and $\sim$600 ps. The periodic nature of these features (t$_2 \approx 2t_1$) typically indicates they correspond to acoustic echoes that come from an acoustic pulse that travels back and forth in the thin film consistently with our previous description of the phenomena in Fig. \ref{fig1}b. 
As seen in the following (Fig. \ref{fig2}), a drastic change of the transient reflectivity signal is revealed when we thermally induce the amorphous to crystalline phase transition. We started with the amorphous GeTe film, and the temperature was increased up to 300 $^\circ$C at a rate of 5 K/s to achieve the amorphous to crystalline transformation. An optical contrast change is observed as expected after the transformation and shown in Fig. \ref{fig2}a. The thermally induced amorphous to crystalline phase transformation was verified in parallel using Raman spectroscopy. The Raman spectra have been recorded with the T64000 Jobin-Yvon spectrometer under microscope with objective $\times$ 50. The laser wavelength was 647.1 nm. The incident power on the sample was 0.2mW. The results are shown in Fig. \ref{fig2}b. The characteristic bands of the amorphous phase are indicated by letters (b) and (c) and are consistent with the literature~\cite{raman,raman2}. When the GeTe is heated up to 300 $^\circ$C (above the crystallization temperature), the characteristic band of the crystalline phase (band (a)) is consistent with prior literature~\cite{raman,raman2}. Note that heating the sample at 400 $^\circ$C makes the band (c) reappear which could correspond to partial melting. The time-resolved signals are shown in Figs. \ref{fig2}(c-d)). Up to 150 $^\circ$C the transient optical reflectivity signal does not evolve significantly as witnessed by the nearly unchanged shape of the acoustic echo (Fig. \ref{fig2}c). Above this temperature the signal drastically evolves up to the studied maximum of 300 $^\circ$C.  
We observe a significant  change of the shape of the acoustic phonon signal with temperature which is also a clear signature of a phase transformation (Figs. \ref{fig2}(c-d)) as it has been already shown for various structural phase transitions~\cite{ruello2009,martensitic}. 
With this measurement, we also evidence a clear evolution of the photoexcited carriers relaxation time discussed in the Supplementary Figure 6~\cite{SM}. Finally, once the crystallization has been thermally achieved above 300 $^\circ$C, we have cooled down the sample and repeated the time-resolved experiments in the range 20-300 $^\circ$C to confirme the stability of the crystalline phase (see Supplementary Figures 6~\cite{SM}). 
Experiments were also conducted at room temperature with both states of GeTe with a pump energy of 3 eV. The transient optical reflectivity signals obtained for both pump energies are shown in Fig. \ref{fig2}(d). The coherent acoustic phonons are shown in Fig. \ref{fig2}(e) once the baseline has been removed and the spectrum of acoustic phonons obtained with a fast Fourier transform (FFT) is presented in Figs. \ref{fig2}(f,g). The signals obtained with a pump photon of 3 eV confirm the drastic evolution of the acoustic pulse shape when the material becomes crystalline. Less pronounced, is the pump photon energy dependence of the acoustic pulse shape either in the amorphous or in the crystalline state. 
The striking information we see on the Fourier components of coherent acoustic phonons signals (Fig. \ref{fig2}(f)) is the red shift of the main component of the spectrum in the case of the crystalline state comparatively to the amorphous state : while centered around  25 GHz for the amorphous state it red shifts down to around 10 GHz for the crystalline state. The sequence of periodic peaks appearing in the spectrum is due to the fact that the FFT is realized with three echoes so that the spectrum is a result of a convolution of the spectrum of a single acoustic echo with the spectrum of a periodic signal associated to the back and forth travel of acoustic phonons. As a comparison, the FFT of a single echo is shown in Fig. \ref{fig2}(g)). The time periodicity of this sequence of peaks in the FFT is given by $T=2H/V$ where $V$ is the sound velocity and $H$ the GeTe thickness. This sequence allows to estimate with a good precision the sound velocity in the amorphous ($V_{a} \approx 2400 m/s$) and in the crystalline ($V_{\alpha} \approx 2500 m/s$) state (see Supplementary Figure 7~\cite{SM}). \\
In the following, we discuss the acoustic echo duration and shape to demonstrate that the signals can be understood only if we consider non-thermal transport of energy achieved by hot carriers in the crystalline phase, due to supersonic electron-hole plasma expansion.

\section{Discussion}
 
\begin{figure*}[t!]
\centerline{\includegraphics[width=12cm]{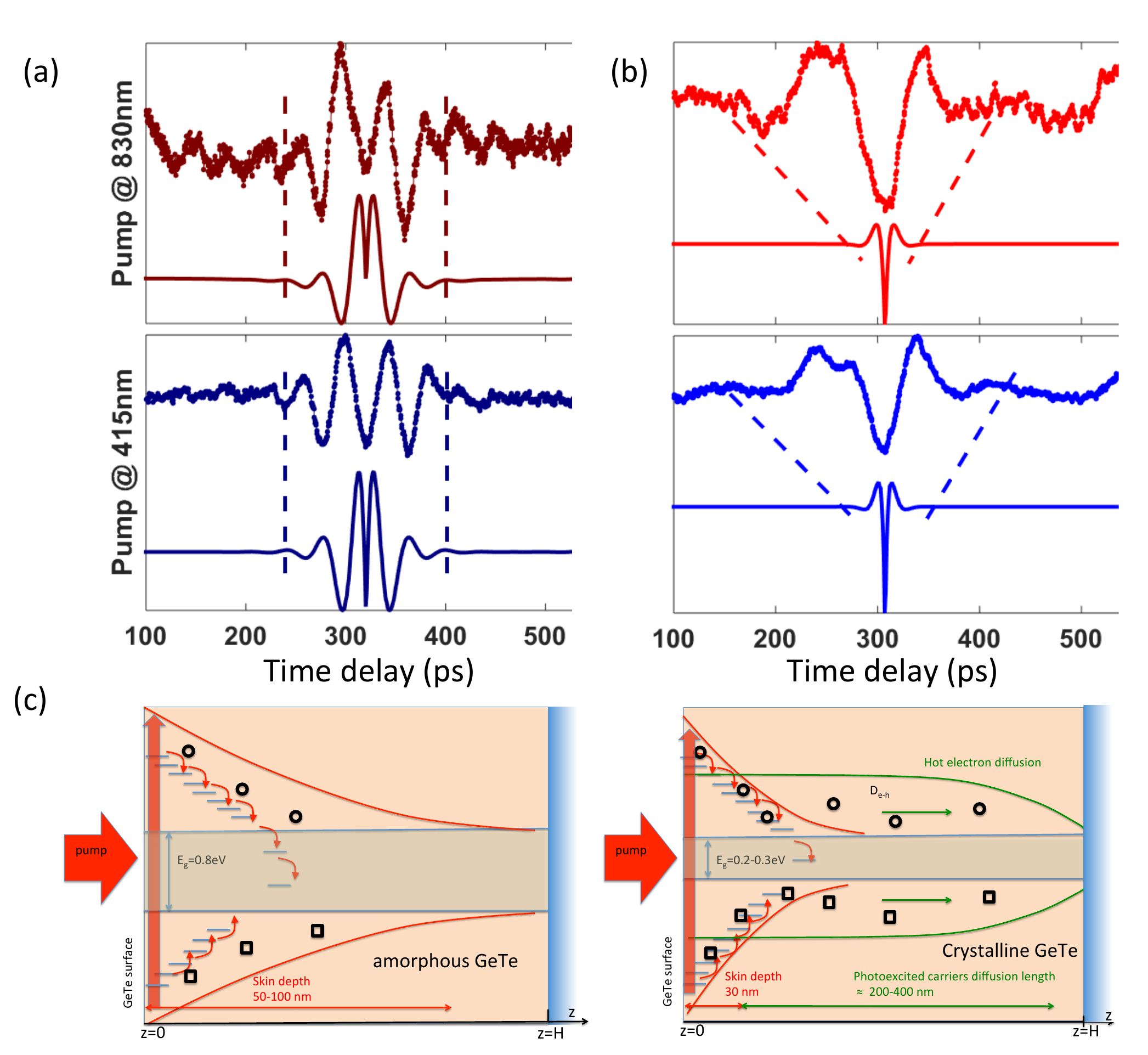}}
\caption{\label{fig3} 
Simulation of the coherent acoustic phonon signals in the amorphous (a) and crystalline (b) states. (c) Sketch of the spatial distribution of the photoexcited carriers in the a-GeTe and $\alpha$-GeTe phase. For the latter one, photoexcited carriers rapidly leave the skin depth which contributes to the strain pulse broadening. Circles and squares are symbols of electrons and holes respectively.}
\end{figure*}

In a semi-infinite solid, when the light impinges on the free surface and leads to a photoinduced stress having the in-depth profile defined by the pump light penetration (i.e. by the skin depth $\xi_{pump}$), the existence of the mechanically free surface leads to emission of an acoustic strain pulse $\eta(z,t)$ having a bipolar shape with  $\eta(z,t)=-\eta_0\times sgn(z-V_{S}t)exp(-\mid{ z-V_{S}t}\mid / \xi)$ (this bipolar pulse is shown in Fig. \ref{fig1}b). 
As discussed in Fig. \ref{fig1}(b), this strain pulse $\eta(z,t)$ travels back and forth in the thin film. This strain field is no directly detected since there is in the time domain a temporal convolution of this strain pulse with the in-depth distribution of the electric field of the probe beam represented by $e^{2ik_{0}\tilde{n}_{GeTe}z}$ in the following equation \cite{tom1,gusev1993,wright}. The theoretical transient optical reflectivity is  $\Delta R/R$=2Re($\delta r/r$) with~\cite{tom1,gusev1993,wright}:

\begin{eqnarray}
\label{drr}
\delta r/r&\simeq&\rho + i\delta\phi\simeq -2ik_{0}\delta z \\
&+& \frac{4ik_{0}\tilde{n}_{GeTe}}{(1-\tilde{n}_{GeTe}^{2})}\frac{d\tilde{n}_{GeTe}}{d\eta}\int_{0}^{\infty}\eta(z,t)e^{2ik_{0}\tilde{n}_{GeTe}z}dz \nonumber.
\end{eqnarray}

where $k_{0}=2\pi/\lambda$ is the probe wave vector in air, $\eta(z,t)$ is the coherent acoustic phonons strain field propagating perpendicularly to the surface of the sample, $\tilde{n}_{GeTe}$ the complex refractive index of GeTe~\cite{wuttig2,optgete} and $d\tilde{n}_{GeTe}/d\eta$ is the photoelastic coefficient. In this model we neglect the very thin transparent layer of ITO, so $z$=0 corresponds to the air/GeTe interface. In Eq. \ref{drr}, the first term corresponds to the contribution of the surface displacement ($\delta z$) of the film at $z$=0. The second term of Eq. \ref{drr} is the photoelastic contribution (due to the modification of the refractive index of the GeTe material induced by the strain field of the coherent acoustic phonons). 
The upper limit of integration has been taken as infinite which is justified since the optical probe penetration is much smaller than the film thickness. The photoelastic coefficient has been estimated according to the $\frac{d\tilde{n}_{GeTe}}{d\eta}=\frac{d\tilde{n}_{GeTe}}{dE}\times \frac{dE}{d\eta}$ where $\frac{d\tilde{n}_{GeTe}}{dE}$ is estimated on the basis of the tabulated data \cite{optgete,wuttig2}  (see supplementary Figure 8~\cite{SM}) and $\frac{dE}{d\eta}$  is the deformation potential parameter. We currently do not know $\frac{dE}{d\eta}$  so in this simulation we can simulate the shape and the acoustic pulse duration of the signal but not the absolute value of magnitude of $\Delta R/R$. 
The calculation is shown in Fig. \ref{fig3}(a). It shows that the detected acoustic phonon signal in the amorphous phase is reasonably reproduced within that model. Our calculation does not perfectly reproduce the shape of the acoustic echo probably because the photoelastic coefficients are slightly different from our estimation. Despite the discrepancies concerning the acoustic echo shape, we are able to see that the pulse duration in the amorphous state is reasonably reproduced. This indicates that the assumption of the photoinduced strain in-depth distribution described by the optical pump penetration is acceptable. Said differently, this strongly suggests a local electron-hole-lattice coupling in the amorphous phase. 
We have carried out the same calculation for the crystalline phase, using the same procedure and employing the optical properties provided by the same Refs.~\cite{optgete,wuttig2} (see Fig. \ref{fig3}(b)).
For this crystalline state, we immediately see the large discrepancy between calculation and experimental measurements, for both the 415 nm and 830 nm pump wavelengths. Due to the closing of the band gap from E$_g \sim 0.8$eV (amorphous) down to E$_g \sim 0.2-0.3$eV (crystalline), the imaginary part of the refractive index significantly increases in the crystalline state. As a consequence, using Eq. \ref{drr}, if the strain profile is solely defined by the in-depth penetration of the pump light, we predict a much shorter pulse duration than the experimental one. 
The fact that the experimental pulse duration is much longer than the calculation (i.e. shift of the phonon spectrum towards the low frequency range), suggests that the strain is induced over a distance exceeding the pump light penetration, i.e. it is non-local in nature. Considering the frequency of the emitted coherent acoustic phonon ($<$10GHz, see Fig. \ref{fig2}(f)), this broadening effect cannot be associated to a rapid heat diffusion out of the skin layer. If light-induced heating effect is responsible, this would mean that the generation of strain is driven by a thermal process, i.e. by the thermoelastic process~\cite{tom1,gusev1993,ruelloLU}.  We remind that such thermoelastic processes, well known in metals~\cite{tom1,gusev1993} is based on a rapid heating of the lattice induced by light absorption. Such heating leads to the lattice expansion and consequently to a strain pulse generation. It is important to know the spatial scale at which this process takes place. For that it is necessary to evaluate the phonon heat flow in $\alpha$-GeTe and to see if the lattice heating takes place only within the optical skin depth of the pump laser or whether the phonon heat transport can occur over a distance larger than the pump skin depth $\xi_{pump}$ hence extending the strain field over the skin depth. Within this thermoelastic process, the typical times for acoustic phonons to leave the region where the pump light has been absorbed is defined by $\tau \sim 2\xi_{pump}/V_S$.  During that time $\tau$, incoherent phonons can diffuse over a typical distance of $L_{th} \sim \sqrt{D_{heat} \tau}$, where $D_{heat}$ is the heat diffusion coefficient. As a consequence, the ratio between the characteristic thermal diffusion length and the optical skin depth of the pump radiation becomes $L_{th}/\xi_{pump}\sim\sqrt{2D_{heat}/(\xi_{pump} V_S)}$. For $\alpha$-GeTe we have, at room temperature $D_{heat}$=2$\times$ 10$^{-6}$ m$^2$.s$^{-1}$ \cite{gonze,mauri} and taking $V_S$ as $\sim$ 2500m/s, we obtain  $L_{th}/\xi_{pump} \sim 0.2<1$. This means that the heat does not have the time to escape the skin depth region characterized by  $\xi_{pump}$ before the longitudinal acoustic phonons leave the same region (we can say that the heat diffusion is subsonic). Thus, the incoherent phonon heat diffusion cannot explain our observed acoustic pulse broadening. \\
Having excluded the thermal effect, it is important to remind that the transport of energy, in particular for photoexcited materials, is not achieved only by incoherent phonons (heat). Some photoexcited carriers can contribute to this transport of energy. Such phenomenon has already been observed in semiconductors GaAs~\cite{wright,kent} and Ge~\cite{chigarev} and was attributed to a supersonic expansion of electron-hole plasma. In that case, the photoexcited carriers rapidly diffuse in the conduction/valence bands and couple to the lattice over a distance larger than the pump skin depth. In the particular case of these semiconductors GaAs and Ge, the coherent acoustic phonons were generated through the electron-hole-acoustic phonon deformation potential coupling mechanism (non-thermal process). This mechanism is possible only if the photoexcited carriers do not recombine before they diffuse out of the skin depth. In this scenario the characteristic acoustic phonon pulsation driven by this rapid plasma expansion is given by $\omega \sim V_S^2/D_{e-h}$ where $D_{e-h}$ (h-holes, e-electron) is the carriers diffusion coefficient~\cite{gusev1993,gusevUps}. In our case, if we assume this process as the driving one, this characteristic frequency is given by the inverse of the duration of the acoustic pulse, i.e. $\omega \sim 1/\Delta t \sim $ 10GHz (that characteristic frequency corresponds to the maximum of the spectrum of a single acoustic echo shown in Fig. \ref{fig2}g). Consequently we deduce from our measurement a photoexcited carrier diffusion coefficient of $D_{e-h}\sim 0.6 \times 10^{-3}$ m$^2$.s$^{-1}$. This value is characteristic for electron and hole diffusion in semiconductors~\cite{adachi}. To show that this value is actually relevant for crystalline GeTe, and if we assume the Einstein relation to be valid, then we can also evaluate the expected diffusion coefficient $D_{e-h}$ in $\alpha$-GeTe from the electrical mobility tabulated in the literature ($\mu_{h,e} \sim$ 100-200 cm$^2$.V$^{-1}$.s$^{-1}$)~\cite{mobi}. (it is worth mentioning that this mobility is 4 orders larger than in the amorphous phase where a typical value $\mu_{h,e} \sim$ 0.1-0.2 cm$^2$.V$^{-1}$.s$^{-1}$ was found~\cite{longeaud}). Using such an approach, we arrive at a value for $D_{e-h}\sim \mu_{h,e} k_B T/q \sim$ 0.2-0.4 $\times$ 10$^{-3}$ m$^2$.s$^{-1}$ at T=300K. Interestingly, this value is consistent with the one deduced from the analysis of diffusive origin of the acoustic echo broadening. Thus, our observation strongly supports that the photoinduced stress is governed by rapid (supersonic) photoexcited carriers in crystalline $\alpha$-GeTe.  A sketch is shown in Fig. \ref{fig3}b where the photoexcited electron-holes are confined within the skin depth for the amorphous phase (and recombine non-radiatively to produce the heat) while the photoexcited carriers rapidly diffuse out of the skin depth with the characteristic diffusion coefficient $D_{e-h}$ in the crystalline state. This supersonic diffusion takes place over a typical distance of $L_D\sim \sqrt{(D_{e-h} \Delta t} \sim$ 200, if we take $D_{e-h} \sim$ 0.6 $\times$ 10$^{-3}$ m$^2$.s$^{-1}$. This means that photoexcited carriers can diffuse over a significant distance and are responsible of a non-thermal transport energy in the crystalline GeTe as we sketched in Figs. \ref{fig1}(a) and \ref{fig3}(b). These results are very interesting and can help us to understand and explain very peculiar out-of-equilibrium properties of photoexcited crystalline GeTe as discussed recently~\cite{kolobov,kolobovPRB,hada}.

\section{Summary} 
Thus, by using a photoacoustic method, we evidence a drastic dependence of the spectrum of the photoinduced strain on the crystalline state of the phase change alloy. The analysis of the photoinduced strain profile provides new insights on the coupling between photoexcited carriers and the lattice. In effect, it remains localized within the photoexcited volume in the amorphous a-GeTe indicating the photoexcited carriers release their energy to the lattice within the volume of light absorption. However, in the crystalline $\alpha$-GeTe phase, the photoinduced strain is not local, i.e. not limited to the volume of light absorption. Surprisingly, a part of the energy is transported over a distance around 7 times larger than the initial penetration depth of the light. Our analysis shows that this effect can be attributed to supersonic diffusion of photoexcited carriers. 
These results demonstrate that the transport of energy is faster than would be if it were solely controlled by heat transfer processes. The existence of such diffusive photoexcited carriers, might have important implications for the ultrafast photoassisted amorphization process whose electronic contributions (i.e. non-thermal processes) versus the thermal effect might play a significant role~\cite{kolobovPRB}.  
These findings could have real implications in the development of sub-THz/THz switching technologies, and in particular for the ultrafast processes for Phase Change Materials based technologies.\\ 

\textbf{Acknowledgements} : 
This project/work is partially funded by the project OptoAc of «Le Mans Acoustique»(R\'egion Pays de la Loire), the European Regional Development Fund (FEDER) and the ANR UP-DOWN (Grant No. ANR-18-CE09-0026-04).
\newpage

\textbf{Supplementray Information: Non-thermal transport of energy driven by photoexcited carriers in switchable solid states of GeTe}  \\

{R. Gu$^{1}$, T. Perrault$^{1}$, V. Juv\'e $^{1}$, G. Vaudel$^{1}$, M. Weis $^{1}$, A. Bulou $^{1}$, N. Chigarev$^{2}$, A. Levchuk$^{1}$, S. Raetz$^{2}$, V. E.  Gusev$^{2}$, Z. Cheng $^{3}$, H. Bhaskaran$^{3}$, P. Ruello$^{1}$}\\

{
$^{1}$Institut des Mol\'ecules et Mat\'eriaux du Mans, UMR 6283 CNRS, Le Mans Universit\'e, 72085 Le Mans,  France\\
$^{2}$Laboratoire d'Acoustique de Le Mans Universit\'e, UMR CNRS 6613, Le Mans Universit\'e, 72085 Le Mans, France\\
$^{3}$ Department of Materials, University of Oxford, United Kingdom.
}

\newpage

\textbf{S1: Determination of the sub-threshold condition to realize picosecond acoustics experiment in a perturbative regime.}\\

Before performing an in situ temperature dependence of the transient reflectivity signal in GeTe, we have determined the pump power threshold that leads to the transformation of the amorphous into the crystalline phase. Since the laser we use has a high repetition rate (80MHz), a heating accumulation effect takes place that can initiate the crystallization. We started the experiment with the amorphous phase and we increased the pump power step by step to determine the threshold above which the amorphous phase is transformed. In Fig. 4(a, b) we show the signal versus the pump fluence. We evidence that only above around 10 mW the signal changes. We have repeated the experiments several times at different places and we have deduced that below 5mW we do not induce the crystallization under the laser action. It is worth mentioning that for high pump power (20mW) the signal looks like the one obtain when we induce a thermodynamic transformation with the in-situ heating system. We can compare the signal obtained when heating up to 300 $^{\circ}$C (Fig. 5(e) in the main manuscript) and the one obtained at 20 mW in Supplementary Figure. 4(b). We observe similar signal even if for the pump-induced crystallization signal, the second echo is less visible than in the thermally-induced amorphization. This is probably due to the fact that the heating induced by the pump beam is not homogeneous in depth. The detailed discussion of this effect is out of the scope of the paper. 
\begin{figure}
\centerline{\includegraphics[width=8cm]{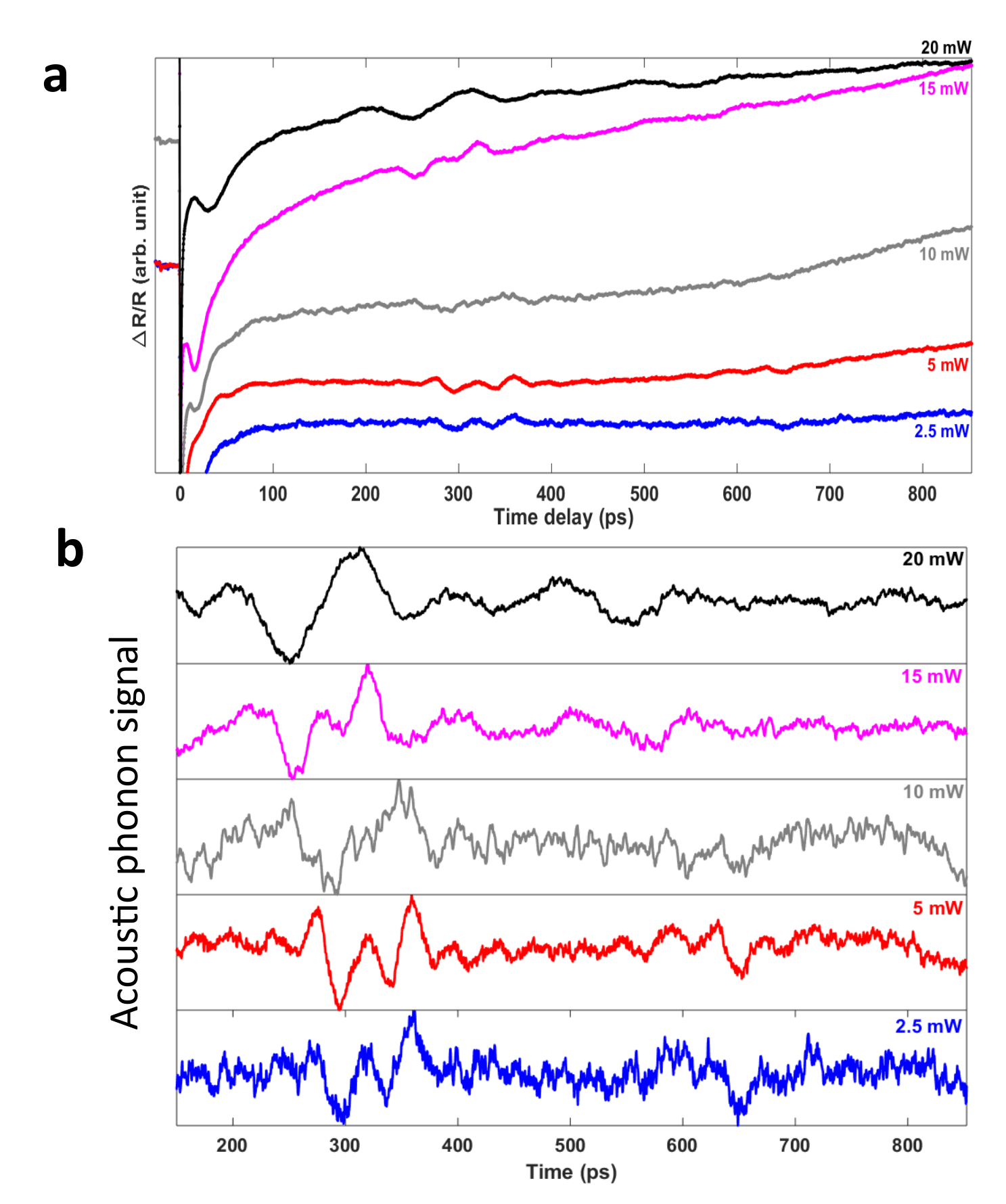}}
\caption{\label{fig4} Pump fluence dependence of the transient reflectivity signal at room temperature. (a) Transient optical reflectivity as a function of the pump fluence. (b) Coherent acoustic phonon signals versus the pump fluence.}
\end{figure}
\begin{figure}
\centerline{\includegraphics[width=7cm]{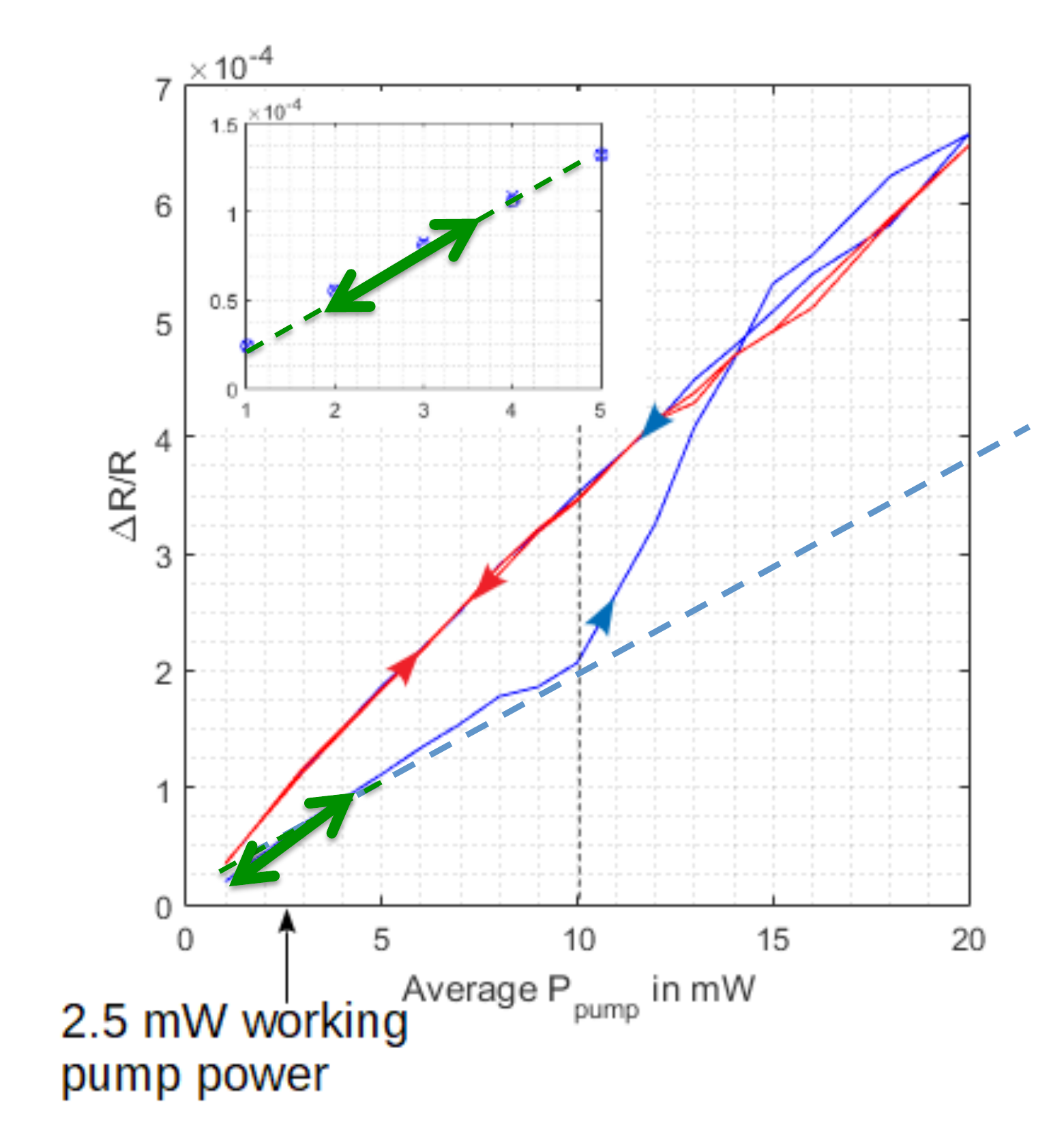}}
\caption{\label{fig5} Sub-threshold optical excitation at room temperature. Transient optical reflectivity signal measured at 12 ns for variable pump fluence. Above 10mW a hysteretic behavior is observed attributed to the transformation of the amorphous state into the crystalline state. Below a pump power of 5mW (60 $\mu$J.cm$^{-2}$), the optical response is reversible.}
\end{figure}
To see if our measurements are done within the perturbative approach, we have also recorded the longest transient optical reflectivity response fixed by our laser repetition rate (80MHz), i.e at 12 ns, versus the pump fluence as shown in Supplementary Figure 5 (room temperature measurements). Starting from the as-grown amorphous phase, when the pump increases we clearly see a non-linear response above typically 10 mW (curve in blue in Supplementary Figure 5) and when the pump power is reduced a clear hysteretic behavior is observed characterizing the transformation of the material under the light beam. A new cycle (red curve) shows after this first ramp, that a linear curve is obtained, with no hysteretic behavior, indicating the material (now in the crystalline state) is stable and the crystalline phase has been obtained by photo-assisted excitation. We did the experiment again, on a different place of the sample, i.e. with an as-grown amorphous layer, but limiting this time the pump power range to [0-5mW] (inset of Supplementary Figure. 5) and we confirmed the stability of the amorphous material under this slight optical excitation (see green curve). It is worth to note that the low fluence threshold we have evidenced in our study is almost 100 times smaller than the fluence threshold used in the recent work of Hase et al. dedicated to picosecond acoustics in Ge$_2$Sb$_2$Te$_5$ \cite{hase}. This difference comes from the high repetition rate laser we used (80MHz) which induces a more pronounced local heat accumulation effect due to the multiple pump excitation that finally drives the crystallization. The thermal properties of the substrate also play a role.\\

\textbf{S2 Ultrafast dynamics}\\

The ultrafast response (fit of the electronic peak decay) reveals a clear change in the carrier-phonon thermalization rate through a change of the carrier relaxation time. The decay of the ultrafast optical reflectivity (figure 6(a)) is modeled with with $\Delta R = A+ Be^{-t/\tau_1}+Ce^{-t/\tau_2}$ where $\tau_1$ and $\tau_2$ are the "fast" and "slow" component.  In figure 6(b), we observe that $\tau_1$ does not significantly change upon crystallization and this could be associated to the cross-correlation decay of pump and probe beam. The time $\tau_2$ exhibits a clear evolution and evidences a longer lifetime in the amorphous state than in the crystalline phase. It is interesting to underline that such variation of hot carriers relaxation time is very similar to the behavior observed in insulator-metal transition \cite{ruello2007} or during the formation of charge-density waves \cite{demsar}. Each time a band gap opened near the Fermi level (i.e. when charge localization occurs) as it does in Refs. \cite{ruello2007,demsar}, a clear slowing down of carrier thermalization has been observed. In the case of GeTe, we do not have a band gap opening, but a band gap increase (from $\sim$0.2-0.3 eV up to 0.8 eV in the amorphous phase). This observation is clear evidence of a dramatic evolution of the photoexcited carrier dynamics due to the phase transformation. 

Once the crystallization has been thermally achieved above 300 $^\circ$C, we have cooled down the sample and repeated the time-resolved experiments in the range 20-300 $^\circ$C. The \textit{in-situ} crystallization of the amorphous GeTe has been evidenced with our pump-probe experiments. We have repeated experiments once we achieved the crystalline phase to demonstrate this state remains stable. We can see in figure 6(c) that the acoustic echo shape remains stable as a function of the temperature which is a direct probe of that stability.\\
\begin{figure*}
\centerline{\includegraphics[width=16cm]{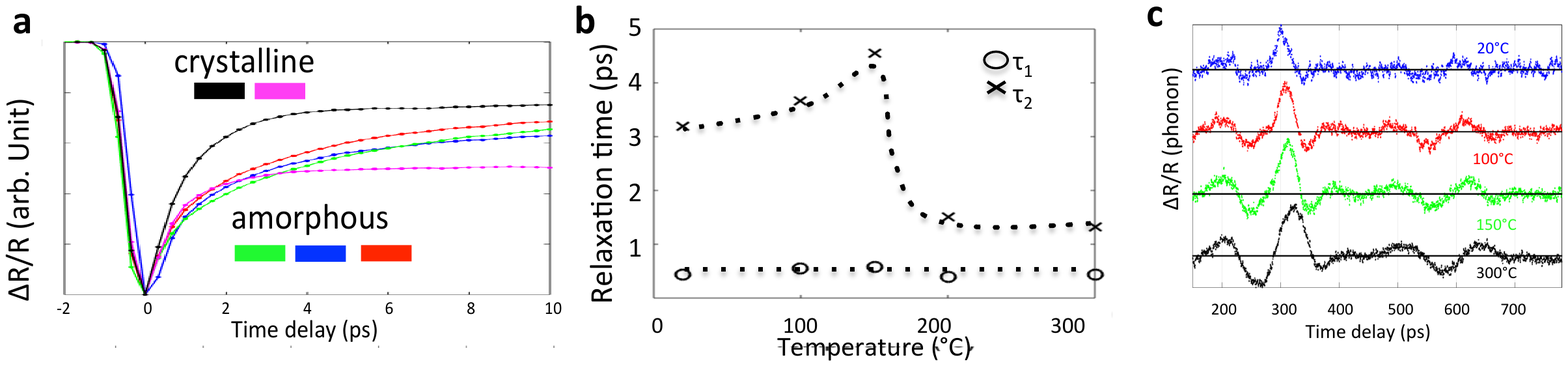}}
\caption{\label{fig6} (a) Temperature dependence of the so-called electronic response. (b) Temperature dependence of the characteristic carrier relaxation time. (c) Acoustic phonon response for variable temperature in the stabilized $\alpha$-GeTe crystalline (ferroelectric) phase. }
\end{figure*}\\
\begin{figure*}
\centerline{\includegraphics[width=16cm]{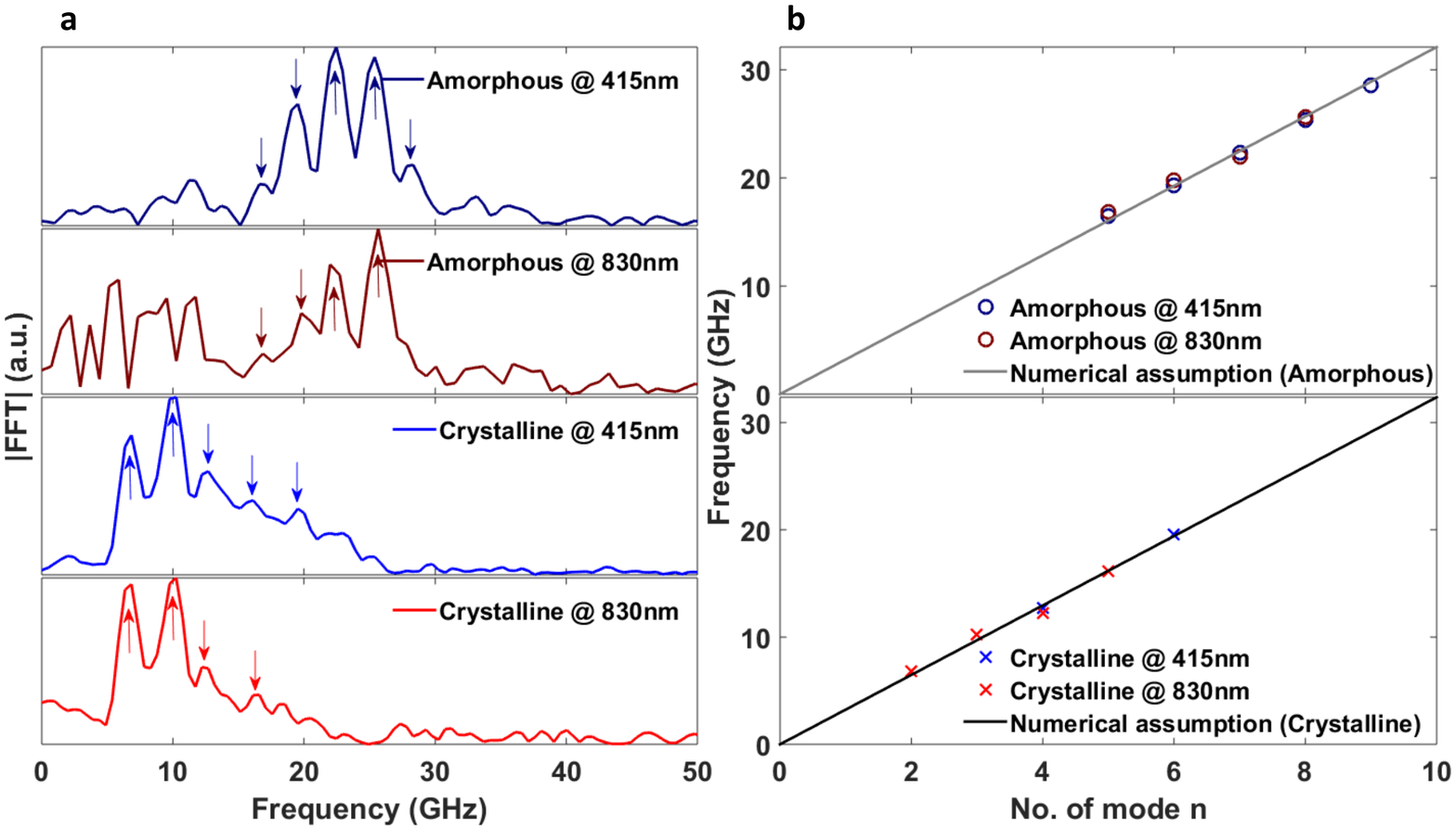}}
\caption{\label{fig8} (a)Spectrum of the time resolved optical reflectivity for the amorphous (top) and crystalline phase (bottom). The sequence of peaks (indicated with arrows) are given by $f_n=nV/2H$, where V is the sound velocity. (b) Evolution of the $f_n$ versus n. The slope of the curves provides the sound velocity.}
\end{figure*}
\textbf{S3 Determination of the sound velocity }\\

The Fast Fourier transform of the entire acoustic phonon signal  is depicted in Fig. 7(a). As mention in the manuscript, the sequence of peaks is de to the presence of several periodic echoes in the signal. This sequence of peaks can reasonably be modeled with the law $f_n=nV/2H$ and the slope of the plot shown in Fig. 7(b) permits to extract the sound velocity.\\

\begin{figure*}
\centerline{\includegraphics[width=12cm]{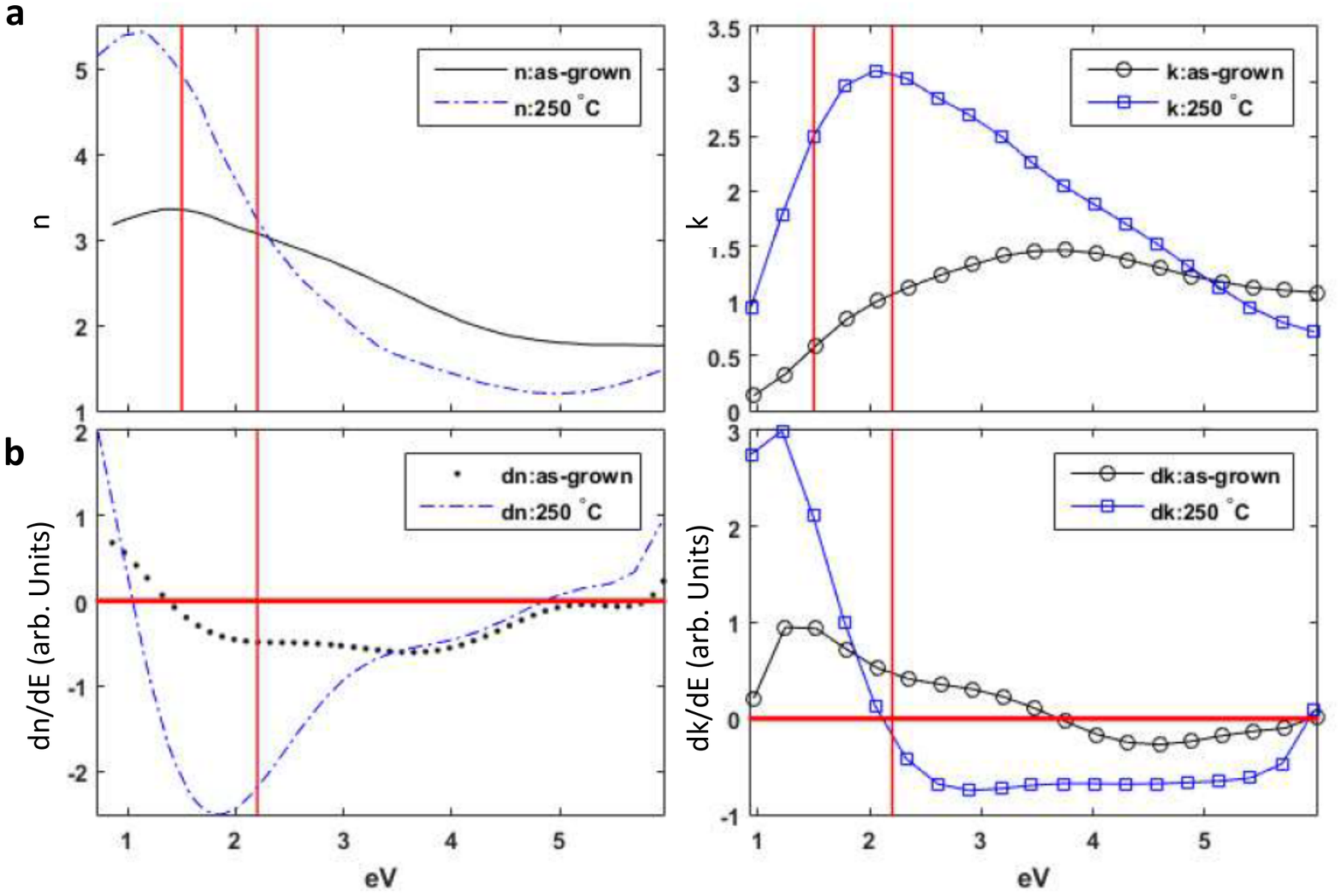}}
\caption{\label{fig7} Qualitative wavelength dependence of the photoelastic coefficient in GeTe. (a) Optical refractive indexes from Ref. (41).  (b) Derivative of the refractive indexes as a function of the photon energy.}
\end{figure*}

\textbf{S4 Determination of the wavelength dependence of the photoelastic coefficients}\\

In the main manuscript we have performed a simulation of the contribution of acoustic phonon response to the transient optical reflectivity signal on the basis of the Thomsen et al model \cite{tom1}. In this model the photoelastic coefficient has been estimated according to the relation  $\frac{d\tilde{n}_{GeTe}}{d\eta}=\frac{d\tilde{n}_{GeTe}}{dE}\times \frac{dE}{d\eta}$ where $\frac{d\tilde{n}_{GeTe}}{dE}$ is estimated on the basis on the tabulated data \cite{wuttig2,optgete}. The term $dE/d\eta$ is the deformation potential parameter that we currently do not know. As a matter of fact, as mentioned in the main manuscript, in this simulation we can reproduce the shape of the signal but not the absolute value of $\Delta R/R$. In Figure. 2 we reproduce the numerical determination of $\frac{d\tilde{n}_{GeTe}}{dE}$ that is obtained by numerical derivative of the refractive index given in Ref. \cite{optgete}. We remind that $\tilde{n}=n+ik$.\\

\end{document}